# CONCEPTUALIZING BLOCKCHAINS: CHARACTERISTICS & APPLICATIONS

Karim Sultan[1], Umar Ruhi[1] and Rubina Lakhani[2]
*[1]Telfer School of Management,*
*[2]Faculty of Engineering,*
*University of Ottawa, Canada*

**ABSTRACT**

Blockchain technology has recently gained widespread attention by media, businesses, public sector agencies, and various international organizations, and it is being regarded as potentially even more disruptive than the Internet. Despite significant interest, there is a dearth of academic literature that describes key components of blockchains and discusses potential applications. This paper aims to address this gap. This paper presents an overview of blockchain technology, identifies the blockchain's key functional characteristics, builds a formal definition, and offers a discussion and classification of current and emerging blockchain applications.

**KEYWORDS**

Blockchain, Cryptocurrency, Bitcoin, Cryptoeconomics, Definition, Business Applications

## 1. INTRODUCTION

The concept of blockchain has become so prevalent in the mainstream, that many are heralding it as the next major disruptive technology. There have been comparisons of its importance to that of the Web and even the Internet (Hernandez, 2017; Mougayar, 2016). While at its core, blockchain is just a method of securely storing and distributing information, it is the potential uses of blockchain technology that make it so empowering: sharing asset transactions between disparate agents with unquestionable transparency – all the while without a controlling central authority. Blockchain creates trust through cryptographic operation by, allowing parties to securely exchange value without the use of an intermediary. Many market sectors are poised for disruption and new startup ventures are vying for dominance in these spaces with a fervor not witnessed since the dot.com boom (Nofer et al., 2017). Despite significant interest, there is a dearth of academic literature that discusses the functional technology underpinning blockchains as well as the potential business applications of this technology. This paper aims to address this gap. This paper is intended to serve as a bridge to blockchain: it provides a technological primer to establish the key concepts and then explores industry applications and trends. We will peek underneath the hood, identify key characteristics of the blockchain technology, and build a formal definition. Next, we will offer a discussion and classification of emerging blockchain applications.

## 2. CONCEPTUAL FOUNDATIONS

### 2.1 Origins & Underpinnings

The innovation of blockchain technology originated from combining the multi-disciplinary fields of software engineering, distributive computing, cryptographic science, and economic game theory. As depicted in Figure 1, blockchains operate at the intersection of these fields that provide the footing for a stable and scalable software infrastructure, a basis for security of digital assets, support for a global decentralized network of peers along with economic incentives for these peers to be good actors in the network. Real-world blockchain applications comprising these multidisciplinary fields are often discussed under the umbrella term of *Cryptoeconomics* – defined as "*a discipline concerned with the production, consumption and transfer of wealth using computer networks, cryptography, and game theory to enhance prosperity of groups in current and future digital market economies*" (Lielacher, 2017).

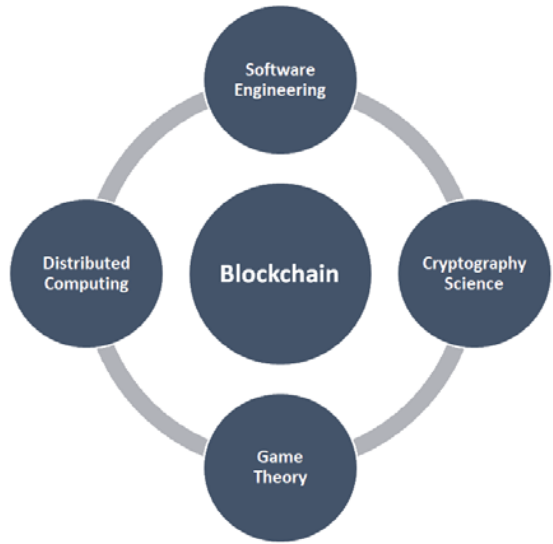

Figure 1. Multidisciplinary Foundations of Blockchain Technology

Blockchain is an underlying construct of Bitcoin, so no discussion of blockchain today is practical without also touching on the Bitcoin cryptocurrency. On October 31, 2008, Satoshi Nakamoto[1] published a brief but groundbreaking paper to a cryptography forum. In it he outlined a way to overcome the double-spend scenario – a problem which plagued previous cryptocurrencies. Despite not mentioning blockchain explicitly, he described its structure as a chain of hashed timestamps: "*Each timestamp includes the previous timestamp in its hash, forming a chain, with each additional timestamp reinforcing the ones behind it*" (Nakamoto, 2008). Although this approach was later refined for Bitcoin, the concept was laid out: a chain of blocks, each cryptographically linked to the previous, using a hash digest. From this we denote that a blockchain is little more than a sequence of records, each hashed and linked to the previous block (Figure 2).

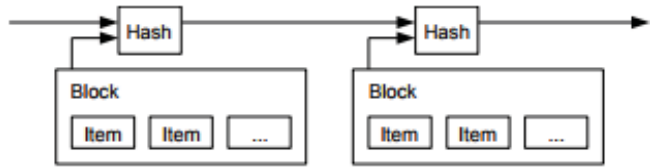

Figure 2. Nakamoto's Blockchain Proposal (Nakamoto, 2008)

However, this approach alone wasn't sufficient to eliminate the double spend issue. In a double spend scenario, the attacker attempts to create a race condition in which they spend the same virtual assets twice before either is validated. For a blockchain to guard against this, Bitcoin needed a method for its network to form consensus. Nakamoto introduced a *Proof-of-Work* model, in which agents would repeatedly hash the block with a random number (nonce) until they achieved a value less than a specified target. Once found, the block would then extend the existing chain. As the blockchain is decentralized over peer-to-peer, any agent can participate in the proof-of-work consensus model to validate transactions (Bitcoin incentives agency with potential economic gains).

---

[1] Satoshi Nakamoto was a pseudonym and has spawned numerous conspiracy theories as to the author's, or authors', actual identity, gender and origin.

The Bitcoin cryptocurrency required a construct into which it could record the order of transactions, verify them, and then secure the entry. Blockchain technology provided this mechanism through a collection of ever extending blocks, where each block represents a pool of transactions, and are cryptographically linked to a parent block via a hash digest. Since the blockchain is represented in a freely distributable file, there is no master copy; it is distributed among all parties. The Bitcoin network relies on this decentralized distribution along with the proof-of-work consensus model to coordinate which blocks are added to the chain and to update all other copies.

The blockchain in Bitcoin functions as a database to store transactions, using a series of inputs and output resembling double entry accounting. Interestingly, bitcoin balances are not maintained, just inputs and outputs. There are no "coins" minted and serialized for consumption, as one may think. By traversing the blockchain, the available balance for a user can be quickly calculated. Any attempt to "corrupt" the blockchain and award more coins in a transaction will fail as the blockchain hashes would have to be recalculated – a computationally hard problem – and the blockchain's decentralized nature ensures other nodes will have legitimate copies.

The first usage of Bitcoin dates to January 3, 2009 at which point Nakamoto created the blockchain genesis block (the first block in the chain and the only one not to have a parent block to link to), and issued himself the first 50 bitcoins (Blockchain Luxembourg S.A., 2017; Zohar, 2015). All blocks in the Bitcoin blockchain trace back to this original transaction. This leads to one more observation: blockchains form an immutable historic record of every transaction from the date of origin in a transparent, decentralized data store.

A blockchain gains its secure, immutable nature by combining two innovations: a cryptographic link between records that makes changes progressively more difficult the longer the chain is, and the distribution of the data to all participating nodes on the decentralized network in which it is expected honest nodes outnumber potential attackers.

## 2.2 Functional Characteristics

Despite being initially linked to Bitcoin, blockchain technology can be used independently in a variety of different use-cases and markets, ranging from insurance to the health industry. A blockchain can be applied in virtually any industry in which assets are managed and transactions occur. It can provide a secure chain of custody for both digital and physical assets through its functional characteristics that facilitate transactions through trust, consensus, security, and smart contracts. These aspects of blockchains are explored in the following sections.

### 2.2.1 Transactions & Smart Contracts

A transaction is an exchange of assets that is managed under the entity service's rules. Such rules are usually operationalized through scripting languages (e.g. Bitcoin's Forth) and are used for advanced transactions (such as escrow and multi-party signatures) to be performed. These rules also form the basis for smart contracts.

A smart contract is a set of logic rules in the form of a coded script which can be embedded into the blockchain to govern a transaction. The contract is executed autonomously and is used to govern the transaction (Buterin, 2016). In this way, contracts act as smart software agents (Stark, 2016). Once a smart contract is embedded in the blockchain, it becomes an autonomous agent that is permanently tamperproof. An application then reads the code when performing a transaction, executes and processes the results.

The contract nature of a smart contract isn't just restricted to application specific code. It can also be used to codify the terms and condition of an agreement into the transaction workflow. Ethereum (the second largest cryptocurrency after Bitcoin) is an alt-coin technology that has been designed to support smart contracts.

### 2.2.2 Consensus & Trust

In events surrounding nuclear disarmament near end of the cold war, President Regan made a Russian proverb famous: "trust, but verify." The same could be claimed for blockchain. It is trusted by consensus as all parties must have identical copies of the blockchain; but each participant is responsible for verifying it.

Blockchain's decentralization is a core strength, as a copy of the database file is owned by all actors. In order to ensure the integrity of each copy, a consensus algorithm is required. The consensus algorithm allows the community to ensure that each added block is legitimate. It also prevents attackers from compromising and forking the chain (Coindesk, 2017). Nakamoto suggested using a proof-of-work approach, in which a hard cryptographic puzzle must be solved by miners (Nakamoto, 2008). Miners expend computing resources and are rewarded for their efforts using various incentives. Other consensus models such as proof-of-stake, proof-of-burn, proof-of-elapsed-time, and proof-of-capacity have been proposed in the literature to overcome some of the weaknesses of the original proof-of-work model by attempting to balance fairness and resource expenditure (Kiayias et al., 2017; Zamfir, 2015).

The nature of blockchain is trustless. Blockchain is designed to eliminate the need for any one entity to gate transactions. It establishes a trust model based on a group consensus, where the network validates transactions and authorizes their addition to the chain. There are no middlemen; the notion of trust becomes implicit as each record in the blockchain is verified by the community which holds multiple copies of the blockchain. By removing trust agents from transactions, blockchain has the ability to disrupt many major industries.

Traditional transaction models rely on central authority to act in the clearinghouse role. Trust is granted to the central authority with an expectation that it will remain honest while verifying and clearing transactions. The instances of records reside with the authority. If the central authority is compromised, either intentionally (manipulation) or unintentionally (hacked), the interlocutor can wreak extensive havoc on the system. The blockchain model eliminates the central authority (Figure 3) by instead disseminating copies of the records to all parties. Each participant maintains their own instance of the blockchain. They broadcast changes by forming new blocks and requesting validation based on the rules of the consensus model. Once validated, the block is added to everyone's chain. The process is potentially safer than the traditional model, and the middleman agent isn't required, invoking a disruption to the status quo.

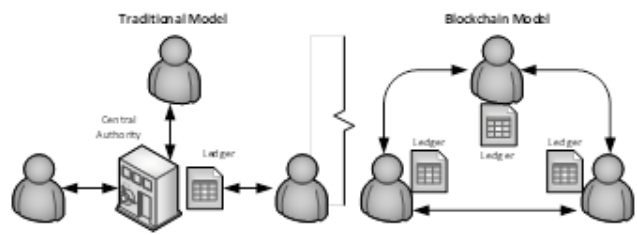

Figure 3. Traditional vs Blockchain Transaction Model

A blockchain relies on mathematics via cryptography to establish independent trust for each transaction, and on computationally expensive consensus models to replace central authorities. Pools of recent transactions are ordered into a block. The block is then cryptographically linked to a chain of blocks (the blockchain) and is verified through a consensus model that involves significant computing resources (mining). As the blockchain is an open-access file replicated on multiple full-nodes of the network, no one entity controls the transaction list. Since each block is hashed and inserted into the chain, it is immutable and serves as a final record of past transactions. An entity cannot change the chain without changing all blocks that follow it, an effort which is computationally hard and expensive. This secures the blockchain and establishes trust independent of a central authority.

### 2.2.3 Public and Private Blockchains

Blockchains can be classified as public, private or hybrid variants, depending on their application (Buterin, 2015; Mougayar, 2016):

- **Public** – Public blockchains have no single owner; are visible by anyone; their consensus process is open to all to participate in; and they are full decentralized. Bitcoin is an example of a public blockchain.

- **Private** – (also called *permissioned)* Private blockchains use privileges to control who can read from and write to the blockchain. Consensus algorithms and mining usually aren't required as a single entity has ownership and controls block creation.

- **Hybrid** – Also known as consortium, these blockchains are public only to a privileged group. The consensus process is controlled by known, privileged servers using a set of rules agreed to by all parties. Copies of the blockchain are only distributed among entitled participants; the network is therefore only partly decentralized.

Although a public blockchain distributes itself in a decentralized peer-to-peer fashion, this isn't necessarily true for a private blockchain. Private blockchains are those used by enterprises to record asset transactions within a limited user base (restricted scope). Hybrid blockchains can be visualized as very small scale public blockchains; they are decentralized only across a limited participant base.

As a summary, we can formalize the afore-noted features of the blockchain into a list of four core characteristics:

- **Immutable** – (permanent and tamper-proof) a blockchain is a permanent record of transactions. Once a block is added, it cannot be altered. This creates trust in the transaction record.
- **Decentralized** – (networked copies) a blockchain is stored in a file that can be accessed and copied by any node on the network. This creates decentralization.
- **Consensus Driven** – (trust verification) each block on the blockchain is verified independently via a Consensus models which provide rules for validating a block, and often use a scarce resource (such as computing power) to show proof that adequate effort was made. In Bitcoin, this is referred to as the mining process. This mechanism works without the use of a central authority or an explicit trust-granting agent.
- **Transparent** – (full transaction history) since the blockchain is an open file, any party can access it and audit transactions[2]. This creates provenance under which asset lifetimes can be tracked.

# 3. TOWARDS A DEFINITION

There are many different blockchain definitions offered, ranging from application-specific to the excessively technical. For example, Coinbase, the world's largest cryptocurrency exchange, defines blockchain as "*a distributed, public ledger that contains the history of every bitcoin transaction*" (Coinbase, 2017). As an application-specific definition, this characterization does not account for the fact that blockchains can be reused for other cryptocurrencies and industry applications independently. The Oxford English Dictionary broadens the definition somewhat, defining blockchain as "*a digital ledger in which transactions made in bitcoin or another cryptocurrency are recorded chronologically and publicly*" (Oxford Dictionaries, 2018). This definition also falls short as blockchain technology can be used independently of Bitcoin and other cryptocurrencies. Both these definitions also highlight the role of a blockchain as a digital ledger, and much of the literature would agree with this. However, this domain is evolving rapidly, and ledger usage is simply a feature of the blockchain but not its essence. This feature only pertains to blockchain applications that focus on managing the exchange of value in the case of virtual assets.

A somewhat broader definition is offered by Webopedia where a blockchain is defined as "*a type of data structure that enables identifying and tracking transactions digitally and sharing this information across a distributed network of computers, creating in a sense a distributed trust network. The distributed ledger technology offered by blockchain provides a transparent and secure means for tracking the ownership and transfer of assets*" (Stroud, 2015). While this definition succeeds in capturing more of the characteristics of blockchain, it highlights distribution as a key feature of blockchain computing without acknowledging that blockchains are not just a distributed technology, but also a decentralized one (Swan, 2015b; Wright and De

---

[2] Websites such as www.blockchain.info provide this service online.

Filippi, 2015). The key distinction here is that while a distributed system partitions work among participants in an optimal fashion, a blockchain requires that each and every participant maintain a full node of the system and enforce its rules independently. In a system where nodes operate on local information (decision locality) to accomplish goals rather than the result of a central ordering influence, this decentralization ensures that pulling the plug is near impossible – only one node needs to remain operational for the network to function.

Evidently, there is a need for a clear and concise definition of blockchain. Based on the theoretical underpinnings of blockchain technology outlined in the previous sections, we provide the following definition of a blockchain:

**"a decentralized database containing sequential, cryptographically linked blocks of digitally signed asset transactions, governed by a consensus model."**

Through this connotative definition, we aim to highlight the core constituents for blockchain technology in that it is a peer-to-peer networked database governed by a set of rules. Furthermore, blockchains represent a shift away from traditional trust agents and a move towards transparency. As a technological building block, it permits applications from a broad swath of industries to take advantage of sharing, tracking, and auditing digital assets. The next section identifies potential opportunities and use-cases for blockchain technology in different industries.

# 4. BLOCKCHAIN BUSINESS APPLICATIONS

## 4.1 Value Proposition of Blockchain Technology

Since the fundamental use-case for blockchain technology is to drive greater transparency and substantiate accuracy of transaction data across the digital information ecosystem, potential applications of such technology are practically endless. In addition to virtual currencies such as Bitcoin, a multitude of other potential blockchain applications and services have been envisioned and discussed by industry pundits and technology research firms (CBInsights, 2018; Mougayar, 2016). To make sense of the blockchain application ecosystem, Mougayar (2016) identifies the value proposition of blockchains as pertaining to the enablement of one or more of six elements denoted by the mnemonic *ATOMIC (Assets, Trust, Ownership, Money, Identity, and Contracts)*. These elements constitute the crux of blockchain business applications by facilitating the creation, management and transfer of digital assets through automated validation rules, cryptographically verified rights and ownerships, and the validation of transactions without requiring intermediaries. Using a blockchain platform implies that each of these six elements is programmable, and by disintermediating these facets, blockchains can enable new services to come to market with cheaper transaction fees and faster execution. In this way, blockchain technology is poised to disrupt many business models which rely on (often costly) intermediaries.

Mougayar (2016) also classifies the role of blockchains as spanning four aspects: blockchain as a development platform; blockchain as a smart contract utility; blockchain as a marketplace, and finally, blockchain as a trusted service application (Rosic, 2017). These potential roles are discussed herewith.

## 4.2 Blockchain as a Development Platform

At present, the development of blockchain applications and services requires a highly specialized skillset, and the state of the blockchain developer toolkits is immature (Aru, 2017). The introduction of *Blockchain as a Service* (BaaS) platforms such as those by Microsoft (Azure) and IBM (Cloud) provide an inexpensive environment for developers to rapidly prototype on test blockchains before deploying to live ones (Sofia, 2016). Other examples in this space include technology platforms that enable secure sharing of data across industrial networks through blockchain's tamperproof ledgers (e.g. Xage[3]); and technologies that offer blockchain enabled verification of data transactions (e.g. Guardtime[4]). These BaaS solutions form the basis for programmable trust, ownership and identity, and also facilitate the operation and governance of enterprise blockchain applications and services.

---

[3] https://xage.com/
[4] https://guardtime.com/

## 4.3 Blockchain as a Smart Contract Utility

Smart contracts provide a programmatic interface to blockchains. Smart contract utility is defined as "*being able to perform useful functions to create, maintain or augment the value of digital assets*" (Sorin et al., 2016). The smart contract, when triggered, transacts value based on digital assets. Utility is captured in code and stored on the blockchain. This code is executed when a predetermined condition occurs. Activities often managed by third party central authorities are mitigated to the blockchain instead, disintermediating transactions (Mougayar, 2015). Examples include escrow, multi-party transactions, digital notarization, and time stamping. Blockchain startups such as R3[5] are collaborating with banks and regulators to operationalize blockchains as a new operating system for financial markets (Irrera, 2017). A specific instance of this functionality can be observed in the example of Visa and DocuSign who have partnered to operate a proof-of-concept blockchain project to streamline vehicle leasing experience for customers by simplifying transaction management between multiple parties including sellers, buyers and insurance companies (Hirson, 2015).

## 4.4 Blockchain as a Marketplace

Any robust ecosystem requires a market for generating value. In the cryptoeconomics marketplace, blockchain provides a payment infrastructure (via cryptocurrencies) and a proof-of-ownership structure (via digital asset tracking). This has enabled peer-to-peer marketplaces with no governing authority, such as OpenBazaar[6] and Soma[7] providing accessible, disintermediated trade (CBInsights, 2018). In these marketplaces, blockchain platforms can be used to directly match buyers and sellers allowing them to transact through smart contracts. Blockchain technology is also being touted as an enabler for next generation online workforce marketplaces in the context of the gig economy that relies on independent contract workers and freelancers for short-term engagements, and the shared economy where consumers increasingly become prosumers. In such instances, blockchain platforms can ensure that service providers are not constrained by any central authority – hence, allowing them to extend flexible offerings, and payment interactions and service transactions can function in a transparent environment (Aitken, 2017).

Overall, as a marketplace enabler, blockchains can be used to operationalize programmable assets, ownership and money.

## 4.5 Trusted-Service Application

Finally, in its role as a trusted service application, blockchain technology comprises end-to-end functionality by facilitating highly specialized applications for any purpose imaginable. This more generalized use of blockchains to enable all types of applications through a combination of programmable assets, trust, ownership, money, identity, and contracts is sometimes referred to as blockchain 2.0 (Bheemaiah, 2016; Swan, 2015a). On the front-end, trusted service applications built on the blockchain using smart contracts can provide disintermediated, secure services to end users. On the backend, many of these applications reside on public blockchains (e.g. Bitcoin and Ethereum) which cannot be shut down or restricted. Furthermore, many companies now provide APIs (application programming interface) allowing developers to build applications using blockchain protocols and mechanisms (Bheemaiah, 2016).

Based on our discussion of potential blockchain business applications, we propose the following 2x2 matrix (

Figure 4) for mapping industry sectors against blockchain scope (public vs private vs hybrid), and blockchain access (as a service or as application). In this context, an application refers to "*a program designed to perform a function or a suite of related functions of benefit to an end user*" (Horak, 2008), while a service is "*a transport of data and/or applications*" (Horak, 2008).

---

[5] https://www.r3.com
[6] https://www.openbazaar.org/
[7] https://soma.co/

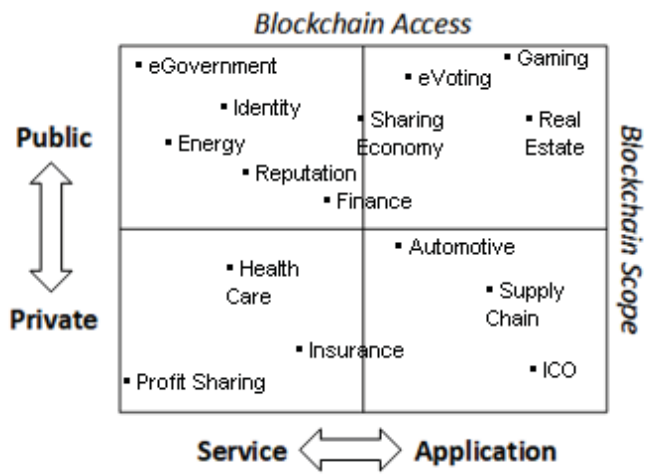

Figure 4. Blockchain Access-vs-Scope Matrix

In order to map horizontal markets on to the above matrix, each offering can be assessed against the following criteria:

- **Access** - Is the primary function to transform data (an application) or to present data (service)?
- **Scope** - Is blockchain usage globally unrestricted (public) or enterprise permissioned (private)?

For contrast, consider the examples of health care and real estate. The health care industry is seeking to facilitate secure transit of patient records via the blockchain. Access is therefore service oriented while the scope remains private to health care partners. In comparison, the real estate industry has shown interest in investigating blockchain for land registration records. This application is meant to be open and transparent to the public.

Overall, these current and envisioned use-cases for blockchain technology are poised to create a global decentralized yet trusted value ecosystem that can lead to exciting new economic opportunities in the public and private sectors alike.

## 5. CONCLUSION

This paper offers a conceptual overview of blockchains through a description of its underlying technological functions and a discussion of its potential business applications. As outlined, contemporary and future blockchain-based innovations span a myriad of use-cases and industries beyond digital currency and the financial sector. Taking this into consideration, we offer a connotative definition that specifies the core elements of blockchain technology independent of Bitcoin. Furthermore, we describe various functional characteristics of blockchain mechanisms, and offer examples of business applications where these mechanisms can potentially be useful.